\documentclass[onecolumn,10pt]{article}
\usepackage{amssymb}
\usepackage{times}

\newcommand{\hil}[1]{\mbox{$\mathcal{#1}$}}

\newcommand{\ket}[1]{| #1 \rangle}

\begin{document}
\date{}
\title{\textbf{Quantum Computation from a \\Quantum Logical Perspective}}
   \vspace{1in}
   
   \author{ Jeffrey Bub\footnote{jbub@umd.edu}\\
\footnotesize Department of Philosophy, University of Maryland,
College Park, MD 20742 \\\footnotesize Perimeter Institute for Theoretical Physics, Waterloo, Canada}
\maketitle

\begin{abstract}It is well-known that Shor's factorization algorithm, Simon's period-finding algorithm, and Deutsch's original XOR algorithm can all be formulated as solutions to a hidden subgroup problem. Here the salient features of the information-processing in the three algorithms are presented from a different perspective, in terms of the way in which the algorithms exploit the non-Boolean quantum logic represented by the projective geometry of Hilbert space. From this quantum logical perspective, the XOR algorithm appears directly as a special case of Simon's algorithm, and all three algorithms can be seen as exploiting the non-Boolean logic represented by the subspace structure of Hilbert space in a similar way. Essentially, a  global property of a function (such as a period, or a disjunctive property) is encoded as a subspace in Hilbert space representing a quantum proposition, which can then be efficiently distinguished from alternative propositions, corresponding to alternative global properties, by a measurement (or sequence of measurements) that identifies the target proposition as the proposition represented by the subspace containing the final state produced by the algorithm.
\end{abstract} 

\section{Introduction}  

Ideally, quantum algorithms allow the computation of certain functions more efficiently than any classical Turing machine. Simon's period-finding algorithm \cite{Simon94,Simon97} achieves an exponential speed-up over any classical algorithm, and Shor's factorization algorithm \cite{Shor94,Shor97}  achieves an exponential speed-up over any known classical algorithm.\footnote{A complete proof of efficiency for Shor's algorithm in the general case was first provided by Ekert and Jozsa \cite{EkertJozsa96}. The algorithm involves a quantum Fourier transform, and the original formulation requires a degree of precision in the implementation of the Fourier transform that is exponential in the size of the input. Barenco et al \cite{Barenco96} showed that an approximate quantum Fourier transform, which does not compromise efficiency, suffices.}

By contrast, Deutsch's original XOR algorithm \cite{Deutsch1985}---the first quantum algorithm with a demonstrated speed-up over over any classical algorithm performing the same computational task---has an even probability of failing, so the improvement in efficiency  is only achieved if the algorithm succeeds. Although a subsequent variation by Cleve \cite{Cleve98} avoids this feature, the speed-up is rather modest: one run of the quantum algorithm versus two runs of a classical algorithm. For Deutsch's problem---a generalization of the XOR problem---the Deutsch-Jozsa algorithm \cite{DeutschJozsa92} achieves a more impressive speed-up: one run of the algorithm versus $2^{n-1} + 1$ runs of a classical deterministic algorithm (for an input  of $n$-bit strings). But even here, a probabilistic classical algorithm yields a solution with high probability after a few runs (see \cite{NielsenChuang}). 

It is well-known that these algorithms can be formulated as solutions to a hidden subgroup problem (see \cite{Jozsa98,Jozsa01}).  Here the salient features of the information-processing in Shor's factorization algorithm, Simon's period-finding algorithm, and Deutsch's original XOR algorithm are presented from a different perspective, in terms of the way in which the algorithms exploit the non-Boolean logic represented by the projective geometry (the subspace structure) of Hilbert space. Essentially, a  global property of a function (such as a period, or a disjunctive property) is encoded as a subspace in Hilbert space representing a quantum proposition, which can then be efficiently distinguished from alternative propositions, corresponding to alternative global properties, by a measurement (or sequence of measurements) that identifies the target proposition as the proposition represented by the subspace containing the final state produced by the algorithm.

Simon's algorithm and Shor's algorithm are algorithms for finding the period of a function. A particular period partitions the domain of the function---the input values for the algorithm---into mutually exclusive and collectively exhaustive subsets. Distinguishing the period from alternative possible periods amounts to distinguishing the corresponding partition from alternative possible partitions. A classical algorithm requires the evaluation of the function for a subset of input values to determine the partition---a computational task that involves an exponentially increasing number of steps as the size of the input increases. The trick in Simon's quantum algorithm, as we will see below, is to represent the alternative possible partitions by Hilbert space subspaces that are orthogonal except for overlaps or intersections. Each subspace is spanned by states of the input register representing the different subsets in the associated partition. A measurement in the computational basis can provide sufficient information to identify the subspace containing the state (after a suitable transformation) and hence the partition associated with the period without evaluating the function at all (in the sense of producing a value in the range of the function for a value in its domain). The algorithm generally has to be run several times because the measurement might be inconclusive, corresponding to an outcome associated with the overlap region, but achieves success in a number of steps that is a polynomial function of the size of the input. In Shor's algorithm, the alternative possible partitions are associated with a family of nested subspaces. The algorithm works as a randomized algorithm  by providing a candidate value for the period in polynomial time, which can be tested in polynomial time. 

At first sight, Deutsch's XOR problem is quite different. It involves the determination of a disjunctive property of a Boolean function. But note that determining the period of a function also amounts to determining a disjunctive property of the function: the disjunction over the different subsets $s_{i}$ of a particular partition of the domain of the function, as opposed to alternative such disjunctions. As we will see below, Deutsch's XOR algorithm works by associating the alternative disjunctions with two Hilbert space planes that are orthogonal except for an intersection in a ray. From this perspective, the XOR algorithm appears directly as a special case of Simon's algorithm, and all three algorithms can be seen as exploiting the non-Boolean logic represented by the projective geometry of Hilbert space in a similar way.

\section{Deutsch's  XOR Algorithm}

In Deutsch's XOR problem \cite{Deutsch1985}, a `black box' or oracle  computes a Boolean function $f: B \rightarrow B$, where $B = \{0,1\}$ is a Boolean algebra (or the additive group of integers mod 2). The problem is to determine whether the function is `constant' (takes the same value for both inputs) or `balanced' (takes a different value for each input). The properties `constant' and `balanced' are two alternative disjunctive properties of the function $f$ (for `constant,' $0 \rightarrow 0$ and $1 \rightarrow 0$ \emph{or} $0 \rightarrow 1$ and $1 \rightarrow 1$; for `balanced,'  $0 \rightarrow 0$ and $1 \rightarrow 1$ \emph{or} $0 \rightarrow 1$ and $1 \rightarrow 0$). Classically, a solution requires two queries to the oracle, for the input values 0 and 1, and a comparison of the outputs. 

Deutsch's algorithm begins by initializing 1-qubit input and output registers  to the state $\ket{0}\ket{0}$ in a standard basis (the computational basis).  A Hadamard transformation---$\ket{0} \rightarrow (\ket{0} + \ket{1})$, $\ket{1} \rightarrow (\ket{0} - \ket{1})$--- is applied to the input register (yielding a linear superposition of states corresponding to the two possible input values 0 and 1) followed by a unitary transformation $U_{f}: \ket{x}\ket{y} \rightarrow \ket{x}\ket{y \oplus f(x)}$ applied to both registers that implements the Boolean function $f$:
\begin{eqnarray}
\ket{0}\ket{0} & \stackrel{H}{\rightarrow} & \frac{1}{\sqrt{2}}(\ket{0} + \ket{1})\ket{0} \\ 
& \stackrel{U_{f}}{\rightarrow} & \frac{1}{\sqrt{2}}(\ket{0}\ket{f(0)} + \ket{1}\ket{f(1)})
\end{eqnarray}

The final composite state of both registers is then one of  two orthogonal states, either (constant):
\begin{eqnarray}
\ket{c_{1}} & = & \frac{1}{\sqrt{2}}(\ket{0}\ket{0} + \ket{1}\ket{0}) \label{eqn:c1} \\
\ket{c_{2}} & = & \frac{1}{\sqrt{2}}(\ket{0}\ket{1} + \ket{1}\ket{1}) \label{eqn:c2}
\end{eqnarray}
or (balanced):
\begin{eqnarray}
\ket{b_{1}} & = & \frac{1}{\sqrt{2}}(\ket{0}\ket{0} + \ket{1}\ket{1}) \label{eqn:b1} \\
\ket{b_{2}} & = & \frac{1}{\sqrt{2}}(\ket{0}\ket{1} + \ket{1}\ket{0}) \label{eqn:b2}
\end{eqnarray}

The states $\ket{c_{1}}, \ket{c_{2}}$ and $\ket{b_{1}}, \ket{b_{2}}$  span two planes $P_{c}, P_{b}$  in 
$\hil{H}^{2}\otimes\hil{H}^{2}$, represented by the projection operators:
\begin{eqnarray}
P_{c} & = & P_{\ket{c_{1}}} + P_{\ket{c_{2}}} \\
P_{b} & = & P_{\ket{b_{1}}} + P_{\ket{b_{2}}}
\end{eqnarray}

Although the states $\ket{c_{1}}, \ket{c_{2}}$ are not orthogonal to the states $\ket{b_{1}}, \ket{b_{2}}$,  the  planes---which represent quantum disjunctions\footnote{Since $P_{\ket{c_{1}}}$ and $P_{\ket{c_{2}}}$ are orthogonal, $P_{c} = P_{\ket{c_{1}}} + P_{\ket{c_{2}}} = P_{\ket{c_{1}}} \vee P_{\ket{c_{2}}}$, where `$\vee$' represents quantum disjunction: the infimum or span (the smallest subspace containing the two component subspaces). Similarly for  $P_{b}$.}---are orthogonal, except for an intersection, so their projection operators commute. The intersection is the line (ray) spanned by the vector:
 \begin{equation}
\frac{1}{2}(\ket{00} + \ket{01} + \ket{10} + \ket{11}) = \frac{1}{\sqrt{2}}(\ket{c_{1}} + \ket{c_{2}}) = \frac{1}{\sqrt{2}}(\ket{b_{1}} + \ket{b_{2}})
\end{equation}

In the `prime' basis spanned by the states $\ket{0'} = H\ket{0}, \ket{1'} = H\ket{1}$, the intersection is the state $\ket{0'}\ket{0'}$, the constant plane is spanned by:
\begin{eqnarray}
\ket{0'}\ket{0'} \\
\ket{0'}\ket{1'} & = & \frac{1}{\sqrt{2}}(\ket{c_{1}} - \ket{c_{2}})
\end{eqnarray} 
and the balanced plane is spanned by:
\begin{eqnarray}
\ket{0'}\ket{0'} \\
\ket{1'}\ket{1'} & = & \frac{1}{\sqrt{2}}(\ket{b_{1}} - \ket{b_{2}})
\end{eqnarray} 
i.e., 
\begin{eqnarray}
P_{c} & = & P_{\ket{0'}\ket{0'}} + P_{\ket{0'}\ket{1'}} \label{eqn:const} \\
P_{b} & = & P_{\ket{0'}\ket{0'}} + P_{\ket{1'}\ket{1'}} \label{eqn:bal}
\end{eqnarray}

To decide whether the function $f$ is constant or balanced we could measure the observable with eigenstates $\ket{0'0'}$, $\ket{0'1'}$, $\ket{1'0'}$, $\ket{1'1'}$ on the final state, which is in the 3-dimensional subspace orthogonal to the vector $\ket{1'0'}$, either in the constant plane or the balanced plane. If the state is in the constant plane,  we will either obtain the outcome $0'0'$ with probability 1/2 (since the final state is at an angle $\pi/4$ to $\ket{0'0'}$), in which case the computation is inconclusive, or the outcome $0'1'$ with probability 1/2. If the state is in the balanced plane, we will either obtain the outcome $0'0'$ with probability 1/2, in which case the computation is inconclusive, or the outcome $1'1'$ with probability 1/2. So in either case, with probability 1/2, we can distinguish whether the function is constant or balanced  in one run of the algorithm by distinguishing between the constant and balanced planes, without evaluating the function at any of its inputs (i.e., without determining in the constant case whether $f$ maps 0 to 0 and 1 to 0, or whether $f$ maps  0 to 1 and 1 to 1, and similarly in the balanced case).\footnote{Equivalently, we could measure the output register. If the outcome is $0'$, the computation is inconclusive. If the outcome is $1'$, we measure the input register. The outcome $1'$ or $0'$ then distinguishes whether the function is constant or balanced.}

Now, a Hadamard transformation applied to the final states of both registers allows the constant  and balanced planes to be distinguished (with probability 1/2) by a measurement in the computational basis. Since $H^{2} = I$, so $\ket{0'0'} \stackrel{H}{\longrightarrow} \ket{00}$, etc.,  a Hadamard transformation of the state amounts to dropping the primes in the representation (\ref{eqn:const}), (\ref{eqn:bal}) for the constant and balanced planes. More precisely, the relationship between the states $\ket{c_{1}}, \ket{c_{2}}, \ket{b_{1}}, \ket{b_{2}}$ in (\ref{eqn:c1}),  (\ref{eqn:c2}),  (\ref{eqn:b1}),  (\ref{eqn:b2}) and the constant and balanced planes defined by $P_{c} = P_{\ket{0'}\ket{0'}} + P_{\ket{0'}\ket{1'}}$ and
$P_{b} = P_{\ket{0'}\ket{0'}} + P_{\ket{1'}\ket{1'}}$ is the same, after the Hadamard transformation of the state, as the relationship between the states $\ket{c_{1}}, \ket{c_{2}}, \ket{b_{1}}, \ket{b_{2}}$ and the planes defined by $P_{c} = P_{\ket{0}\ket{0}} + P_{\ket{0}\ket{1}}$ and
$P_{b} = P_{\ket{0}\ket{0}} + P_{\ket{1}\ket{1}}$. That is, under the Hadamard transformation: 
\begin{eqnarray}
\ket{c_{1}} & \rightarrow & \frac{1}{\sqrt{2}}(\ket{0}\ket{0} + \ket{0}\ket{1})  \\
\ket{c_{2}} & \rightarrow & \frac{1}{\sqrt{2}}(\ket{0}\ket{0} - \ket{0}\ket{1}) 
\end{eqnarray}
and:
\begin{eqnarray}
\ket{b_{1}} & \rightarrow & \frac{1}{\sqrt{2}}(\ket{0}\ket{0} + \ket{1}\ket{1})  \\
\ket{b_{2}} & \rightarrow & \frac{1}{\sqrt{2}}(\ket{0}\ket{0} - \ket{1}\ket{1}) 
\end{eqnarray}
So the transformed constant plane $HP_{c}$ is spanned by:
\begin{eqnarray}
\ket{0}\ket{0} & = & \frac{1}{\sqrt{2}}(H\ket{c_{1}} + H\ket{c_{2}}) \\
\ket{0}\ket{1} & = & \frac{1}{\sqrt{2}}(H\ket{c_{1}} - H\ket{c_{2}})
\end{eqnarray} 
and the transformed balanced plane $HP_{b}$ is spanned by:
\begin{eqnarray}
\ket{0}\ket{0} & = & \frac{1}{\sqrt{2}}(H\ket{b_{1}} + H\ket{b_{2}}) \\
\ket{1}\ket{1} & = & \frac{1}{\sqrt{2}}(H\ket{b_{1}} - H\ket{b_{2}})
\end{eqnarray} 

This  is crucial for an evaluation of the efficiency of the algorithm relative to a classical algorithm. The number of relevant computational steps in a quantum algorithm is conventionally counted as the number of applications of unitary transformations and measurements required to yield a solution, where the unitary transformations belong to a standard set of elementary unitary gates that form a universal set, and the measurements are in the computational basis. 

In Cleve's variation, the two registers are initialized to $\ket{0}$ and $\ket{1}$, respectively (instead of to $\ket{0}$ and $\ket{0}$). A Hadamard transformation to both registers  yields the transition:
\begin{eqnarray}
\ket{0}\ket{1} \stackrel{H}{\rightarrow} \frac{\ket{0} + \ket{1}}{\sqrt{2}}\frac{\ket{0} - \ket{1}}{\sqrt{2}} \label{eqn:init}
\end{eqnarray}
Since
\begin{equation}
U_{f}\ket{x}\ket{y} = \ket{x}\ket{y\oplus f(x)}
\end{equation}
it follows that 
\begin{equation}
U_{f}\ket{x}\frac{\ket{0} - \ket{1}}{\sqrt{2}} = \left \{\begin{array}{c}
 \ket{x}\frac{\ket{0} - \ket{1}}{\sqrt{2}} \mbox{ if $f(x) = 0$} \vspace{.1 in} \\
 - \ket{x}\frac{\ket{0} - \ket{1}}{\sqrt{2}} \mbox{ if $f(x) = 1$}
 \end{array} \right.
\end{equation}
which can be expressed as
\begin{equation}
U_{f}\ket{x}\frac{\ket{0} - \ket{1}}{\sqrt{2}}  = (-1)^{f(x)}\ket{x}\frac{\ket{0} - \ket{1}}{\sqrt{2}} \label{eqn:phase}
\end{equation}
The value of the function now appears as a phase of the final state of the input register.  For the input state $1/\sqrt{2}(\ket{0} + \ket{1})$, we have:
\begin{equation}
U_{f}\frac{\ket{0} + \ket{1}}{\sqrt{2}}\frac{\ket{0} - \ket{1}}{\sqrt{2}}  = 
\frac{(-1)^{f(0)}\ket{0} + (-1)^{f(1)}\ket{1}}{\sqrt{2}} \frac{\ket{0} - \ket{1}}{\sqrt{2}} 
\end{equation}
which can be expressed as:
\begin{equation}
U_{f}\frac{\ket{0} + \ket{1}}{\sqrt{2}}\frac{\ket{0} - \ket{1}}{\sqrt{2}}  = \left \{\begin{array}{c}
 \pm \frac{\ket{0} + \ket{1}}{\sqrt{2}}\frac{\ket{0} - \ket{1}}{\sqrt{2}} = \pm \ket{0'}\ket{1'} \mbox{ if $f(0) = f(1)$} \vspace{.1 in} \\
  \pm \frac{\ket{0} - \ket{1}}{\sqrt{2}}\frac{\ket{0} - \ket{1}}{\sqrt{2}} = \pm\ket{1'}\ket{1'}  \mbox{ if $f(0) \neq f(1)$} 
  \end{array} \right.
\end{equation}

Instead of the final state of the two registers ending up as one of two orthogonal states in the constant plane, or as one of two orthogonal states in the balanced plane, the final state now ends up as $\pm\ket{0'1'}$ in the constant plane, or as $\pm\ket{1'1'}$ in the balanced plane, and these states can be distinguished because they are orthogonal. So we can decide with certainty whether the function is constant or balanced after only one run of the algorithm. In fact, we can distinguish these two possibilities by simply measuring the input register in the prime basis, and since  a final Hadamard transformation on the state of the input register  takes $\ket{0'}$ to $\ket{0}$ and $\ket{1'}$ to 
$\ket{1}$), we can distinguish the two planes by measuring the input register in the computational basis. Note that the state of the output register is unchanged: at the end of the process it is in the  state $\ket{1'} = H\ket{1}$ as in (\ref{eqn:init}). 

Deutsch's XOR problem can be generalized to the problem (`Deutsch's problem') of determining whether a Boolean function $f:B^{n} \rightarrow B$ is constant or whether it is balanced, where it is promised that the function is either constant or balanced.  `Balanced' here means that the function takes the values 0 and 1 an equal number of times, i.e., $2^{n-1}$ times each. Exploiting the Cleve variation of the XOR algorithm, the Deutsch-Jozsa algorithm \cite{DeutschJozsa92} determines whether $f$ is constant or balanced in one run.

The algorithm proceeds by setting the input $n$-qubit register to the state $\ket{0}$ (an abbreviation for the state $\ket{0 \cdots 0} = \ket{0} \cdots \ket{0}$) and the output 1-qubit register to the state $\ket{1}$, as in the Cleve variation of the XOR algorithm. An $n$-fold Hadamard transformation is applied to the input register and a Hadamard transformation to the output register, followed by the unitary transformation $U_{f}$ to both registers, and finally an $n$-fold Hadamard transformation to the input register.

The state of the input register ends up as:
\begin{equation}
\sum_{y}\sum_{x} \frac{(-1)^{x\cdot y + f(x)}}{2^{n}} \ket{y} = \sum_{x}\frac{(-1)^{f(x)}}{2^{n}} \ket{0 \ldots 0} + \ldots \label{eqn:DJ}
\end{equation}
The amplitude of the state $\ket{0 \ldots 0}$ in the linear superposition (\ref{eqn:DJ}) is $\sum_{x}\frac{(-1)^{f(x)}}{2^{n}}$. If $f$ is constant, this coefficient is $\pm 1$, so the coefficients of the other terms must all be 0. If $f$ is balanced, $f(x) = 0$ for half the values of $x$ and $f(x) = 1$ for the other half, so the positive and negative contributions to the coefficient of $\ket{0 \ldots 0}$ cancel to 0. In other words, if $f$ is constant, the state of the input register is $\pm\ket{0 \ldots 0}$; if $f$ is balanced, the state is in the orthogonal subspace. Since the initial and final Hadamard transformations can be implemented efficiently, i.e., with a number of elementary unitary gates that is only a polynomial function of the size of the input, the algorithm is exponentially faster than any classical algorithm.

This is the usual way of describing how the algorithm works. To see what is going on from a quantum logical perspective, in terms of the subspaces representing the constant and balanced quantum propositions,  consider, for simplicity, the case $n = 2$. After the transformation $U_{f}$, but before the final Hadamard transformation, the state of the input register is either (constant):
\begin{equation}
\pm\frac{1}{2}(\ket{00} + \ket{01} + \ket{10} + \ket{11})
\end{equation}
or (balanced):
\begin{equation}
\frac{1}{2}(\pm\ket{00} \pm \ket{01} \pm \ket{10} \pm \ket{11})
\end{equation}
where two of the coefficients are $+1$ and two of the coefficients are $-1$. Evidently, there are three (distinct, up to an overall phase $e^{i\pi} = -1$) mutually orthogonal such balanced states, and they are all orthogonal to the constant state. So the three balanced states lie in a 3-dimensional subspace orthogonal to the constant state and can therefore be distinguished from the constant state. The final Hadamard transformation transforms the constant state to the state to $\ket{00}$:
\begin{equation}
\frac{1}{2}(\ket{00} + \ket{01} + \ket{10} + \ket{11}) \stackrel{H}{\longrightarrow} \ket{00}
\end{equation}
 and the three balanced states to states in the 3-dimensional subspace orthogonal to $\ket{00}$. So to decide whether the function is constant or balanced we need only measure the input register in the computational basis and check whether it is in the state $\ket{00}$.
 
 The Cleve variation of Deutsch's XOR algorithm and its application to the Deutsch-Jozsa algorithm for Deutsch's problem are included here for completeness in illustrating the geometric features of these algorithms. The relevant feature of the information-processing for comparison with Simon's algorithm and Shor's algorithm is already, and more clearly shown, in the original XOR algorithm, as we will see below.

\section{Simon's Algorithm}

Simon's problem is to find the period $r$ of a periodic Boolean function $f:B^{n} \rightarrow B^{n}$, i.e., a 
function for which
\begin{equation}
f(x_{i}) = f(x_{j}) \mbox{ if and only if $x_{j} = x_{i} \oplus r$, for all $x_{i},x_{j} \in B^{n}$}
\end{equation}
Note that since $x\oplus r \oplus r =  x$, the function is 2-to-1.

Since $f$ is periodic, the possible outputs of $f$---the values of $f$ for the different inputs---partition the set of input values into mutually exclusive and collectively exhaustive subsets, and these subsets depend on the period. So, determining the period of $f$ amounts to distinguishing the partition corresponding to the period from alternative partitions corresponding to alternative possible periods.

Simon's algorithm solves the problem efficiently, with an exponential speed-up over any  classical algorithm (see \cite{Simon94,Simon97}). The input and output registers are initialized to the  state $\ket{0}\ket{0}$ in the computational basis (where, as before, $\ket{0}$ is an abbreviation for $\ket{0} \ldots \ket{0} = \ket{0 \ldots 0}$) and the state is evolved as follows:
\begin{eqnarray}
\ket{0}\ket{0} & \stackrel{H}{\longrightarrow} & \frac{1}{\sqrt{2^{n}}} \sum_{x=0}^{2^{n}-1}\ket{x}\ket{0} \\
& \stackrel{U_{f}}{\longrightarrow} &  \frac{1}{\sqrt{2^{n}}} \sum_{x}\ket{x}\ket{f(x)} \\
& & = \frac{1}{\sqrt{2^{n-1}}}\sum_{x_{i}}\frac{\ket{x_{i}} + \ket{x_{i} \oplus r}}{\sqrt{2}} \ket{f(x_{i})}
\end{eqnarray}
where $U_{f}$ is the unitary transformation implementing the Boolean function as:
\begin{equation}
U_{f}: \ket{x}\ket{y} \rightarrow \ket{x}\ket{y \oplus f(x)}
\end{equation}

A measurement of the output register would leave  the input register in a state of the form:\footnote{Considering a measurement of the output register here is simply a pedagogical device, for clarity. Instead, we could refer to the reduced state of the input register, which is a mixture of states of the form (\ref{eqn:offset}). No actual measurement of the output register is required, only a measurement of the input register.}
\begin{equation}
\frac{\ket{x_{i}} + \ket{x_{i}\oplus r}}{\sqrt{2}} \label{eqn:offset}
\end{equation}
This state contains the information $r$, but summed with an unwanted randomly chosen offset $x_{i}$ that depends on the measurement outcome. A direct measurement of the state label would  yield any $x \in B^{n}$ equiprobably, providing no information about $r$.

The application of a final Hadamard transformation yields:
\begin{eqnarray}
\frac{\ket{x_{i}} + \ket{x_{i}\oplus r}}{\sqrt{2}} & \stackrel{H}{\longrightarrow} & \frac{1}{\sqrt{2^{n}}}\sum_{y \in B^{n}}\frac{(-1)^{x_{i}\cdot y} + (-1)^{(x_{i}\oplus r)\cdot y}}{\sqrt{2}}\ket{y} \\
& = & \sum_{y:r\cdot y = 0} \frac{(-1)^{x_{i}\cdot y}}{\sqrt{2^{n-1}}}\ket{y} \label{eqn:SimonHadamard}
\end{eqnarray}
where the last equality follows  because terms interfere destructively if $r\cdot y =1$. A measurement of  the input register in the computational basis yields a value $y$ (equiprobably) such that $r\cdot y = 0$. Repeating the algorithm sufficiently many times yields enough values $y_{i}$ so that $r$ can be determined by solving the linear equations $r\cdot y_{1} = 0, \ldots, r\cdot y_{k} = 0$.

To see how the algorithm works quantum logically in terms of the subspaces representing the relevant quantum propositions, consider  the case $n = 2$. There are $2^{2} - 1 = 3$ possible values of the period $r$: 01, 10, 11, and the corresponding partitions are:
\begin{description}\centering
\item[$r = 01:$] $\{00,01\}, \{10,11\}$
\item[$r = 10:$] $\{00,10\}, \{01,11\}$
\item[$r = 11:$] $\{00,11\}, \{01,10\}$
\end{description}
The corresponding states of the input and output registers after the unitary transformation $U_{f}$ are:
\begin{description}\centering
\item[$r = 01:$] $\frac{1}{2}(\ket{00} + \ket{01})\ket{f(00)} + \frac{1}{2}(\ket{10} + \ket{11})\ket{f(10)}$
\item[$r = 10:$] $\frac{1}{2}(\ket{00} + \ket{10})\ket{f(00)} + \frac{1}{2}(\ket{01} + \ket{11})\ket{f(01)}$
\item[$r = 11:$] $\frac{1}{2}(\ket{00} + \ket{11})\ket{f(00)} + \frac{1}{2}(\ket{01} + \ket{10})\ket{f(01)}$
\end{description}

Notice that this case reduces to the same construction as in Deutsch's XOR algorithm. For $r = 10$ the input register states are:
\begin{eqnarray}
\ket{c_{1}}  & = & \frac{1}{\sqrt{2}}(\ket{00} + \ket{10}) \\
\ket{c_{2}} & = & \frac{1}{\sqrt{2}}(\ket{01} + \ket{11})
\end{eqnarray}
and for $r = 11$ the input register states are:
\begin{eqnarray}
\ket{b_{1}} & = & \frac{1}{\sqrt{2}}(\ket{00} + \ket{11}) \\
\ket{b_{2}} & = & \frac{1}{\sqrt{2}}(\ket{01} + \ket{10})
\end{eqnarray}
depending on the outcome of a measurement of the output register. Here the orthogonal states $\ket{c_{1}}, \ket{c_{2}}$ represent the two subsets of the partition associated with the period $r = 10$, the orthogonal states $\ket{b_{1}}, \ket{b_{2}}$ represent the two subsets of the partition associated with the period  $r = 11$, and the orthogonal states $\ket{00} + \ket{01}, \ket{10} + \ket{11}$ represent the two subsets of the partition associated with the period $r = 01$.  

The three partitions  associated with the three possible periods are represented by three planes in $\hil{H}^{2}\otimes \hil{H}^{2}$, which correspond to the constant and balanced planes in Deutsch's XOR algorithm, and a third orthogonal plane. While the states representing subsets of different partitions associated with different periods are nonorthogonal, the three planes (spanned by these states) are mutually orthogonal, except for an intersection in the ray spanned by the vector $\ket{0'0'}$ in the prime basis (i.e., their projection operators commute):
\begin{description}\centering
\item[$r = 01:$] plane spanned by $\ket{0'0'}, \ket{1'0'}$
\item[$r = 10:$] plane spanned by $\ket{0'0'}, \ket{0'1'}$ (corresponds to `constant'  plane)
\item[$r = 11:$] plane spanned by $\ket{0'0'}, \ket{1'1'}$ (corresponds to `balanced' plane)
\end{description}

We cannot identify the period by a measurement that identifies the state of the input register as a state representing a particular subset of a particular partition, because the states representing subsets of different partitions associated with different periods are non-orthogonal. We could identify the plane corresponding to the period by measuring the input register in the prime basis, but---as in Deutsch's XOR algorithm---the final Hadamard transformation (which, as we have seen, amounts to dropping the primes: $\ket{0'0'} \stackrel{H}{\rightarrow} \ket{00}$, etc.) allows the plane corresponding to the period to be measured in the computational basis. The three possible periods can therefore be distinguished by measuring the observable with eigenstates $\ket{00}, \ket{01}, \ket{10}, \ket{11}$, except when the state of the register is projected by the measurement  onto the state $\ket{00}$ (which occurs with probability 1/2). So the algorithm will generally have to be repeated until we find an outcome that is not 00.

The $n=2$ case of Simon's algorithm essentially reduces to Deutsch's XOR algorithm. In the $n = 3$ case (which suffices to exhibit the general feature of the algorithm) there are $2^{3} -1 = 7$ possible periods: 001, 010, 011, 100, 101, 110, 111. For the period $r = 001$, the state of the two registers after the unitary transformation $U_{f}$ is:
\begin{eqnarray}
\lefteqn{\frac{1}{2\sqrt{2}}(\ket{000} + \ket{001})\ket{f(000)} + \frac{1}{2\sqrt{2}}(\ket{010} + \ket{011})\ket{f(010)}} \nonumber \\
& & + \frac{1}{2\sqrt{2}}(\ket{100} + \ket{101})\ket{f(100)} + \frac{1}{2\sqrt{2}}(\ket{110} + \ket{111})\ket{f(110)}
\end{eqnarray}
A measurement of the output register would leave the input register in one of four states, depending on the outcome of the measurement:
\begin{eqnarray*}
\frac{1}{\sqrt{2}}(\ket{000} + \ket {001}) & = & \frac{1}{2}(\ket{0'0'0'} + \ket{0'1'0'} + \ket{1'0'0'} + \ket{1'1'0'})\\
\frac{1}{\sqrt{2}}(\ket{010} + \ket {011}) & = & \frac{1}{2}(\ket{0'0'0'} - \ket{0'1'0'} + \ket{1'0'0'} - \ket{1'1'0'})\\
\frac{1}{\sqrt{2}}(\ket{100} + \ket {101}) & = & \frac{1}{2}(\ket{0'0'0'} + \ket{0'1'0'} - \ket{1'0'0'} - \ket{1'1'0'}) \\
\frac{1}{\sqrt{2}}(\ket{110} + \ket {111}) & = & \frac{1}{2}(\ket{0'0'0'} - \ket{0'1'0'} - \ket{1'0'0'} + \ket{1'1'0'})
\end{eqnarray*}
Applying a Hadamard transformation amounts to dropping the primes. So if the period is $r = 001$,  the state of the input register ends up in the 4-dimensional subspace of $\hil{H}^{2} \otimes \hil{H}^{2} \otimes \hil{H}^{2}$ spanned by the vectors: $\ket{000}, \ket{010}, \ket{100}, \ket{110}$. 

A similar analysis applies to the other six possible periods. The corresponding subspaces are spanned by the following vectors:
\begin{description}\centering
\item{r = 001:} $\mbox{ } \ket{000}, \ket{010}, \ket{100}, \ket{110}$
\item{r = 010:} $\mbox{ } \ket{000}, \ket{001}, \ket{100}, \ket{101}$
\item{r = 011:} $\mbox{ } \ket{000}, \ket{011}, \ket{100}, \ket{111}$
\item{r = 100:} $\mbox{ } \ket{000}, \ket{001}, \ket{010}, \ket{011}$
\item{r = 101:} $\mbox{ } \ket{000}, \ket{010}, \ket{101}, \ket{111}$
\item{r = 110:} $\mbox{ } \ket{000}, \ket{001}, \ket{110}, \ket{111}$
\item{r = 111:} $\mbox{ } \ket{000}, \ket{011}, \ket{101}, \ket{110}$
\end{description}

These subspaces are orthogonal except for intersections in 2-dimensional planes. The period  can be found by measuring in the computational basis. Repetitions of the measurement will eventually yield sufficiently many distinct values to determine the subspace containing the final state. In this case, it is clear by examining the above list that two values distinct from 000 suffice to determine the subspace, and these are just the values $y_{i}$ for which $y_{i} \cdot  r =  0$.

\section{Shor's Algorithm}

Shor's factorization algorithm exploits the fact that the two prime factors $p, q$ of a positive integer $N = pq$ can be found by determining the period of a function 
$f(x) = a^{x}\mbox{ mod $N$}$, for any $a < N$  which is coprime to $N$, i.e., has no common factors with $N$ (other than 1).  The period $r$ of $f(x)$ depends on $a$ and $N$. Once we know the period, we can factor $N$ if $r$ is even and $a^{r/2} \neq -1 \mbox{ mod $N$}$, which will be the case with probability greater than 1/2 if $a$ is chosen randomly. (If not, we choose another value of $a$.) The factors of $N$ are the greatest common factors of $a^{r/2} \pm 1$ and $N$, which can be found in polynomial time by the Euclidean algorithm. (For these number-theoretic results, see \cite[Appendix 4]{NielsenChuang}.) So the problem of factorizing a composite integer $N$ that is the product of two primes reduces to the problem of finding the period of a certain function $f: Z_{s} \rightarrow Z_{N}$, where $Z_{n}$ is the additive group of integers mod $n$ (rather than $B^{n}$, the $n$-fold Cartesian product of a Boolean algebra $B$, as in Simon's algorithm).

Note that $f(x+r) = f(x)$ if $x+r \leq s$. The function $f$ is periodic if $r$ divides $s$ exactly, otherwise it is almost periodic. 

Consider first the general form of the algorithm, as it is usually formulated. The input register ($k$ qubits, whose states are represented on an $s$-dimensional Hilbert space $\hil{H}^{s}$, where $s = 2^{k}$) is initialized to the state $\ket{0} \in \hil{H}^{s}$ and the output register  to the state $\ket{0} \in \hil{H}^{N}$. A $k$-fold Hadamard transformation is applied to the input register, followed by the unitary transformation $U_{f}$ which implements the function $f(x) = a^{x}\mbox{ mod $N$}$:
\begin{eqnarray}
\ket{0}\ket{0} & \stackrel{H}{\longrightarrow} & \frac{1}{\sqrt{s}}\sum_{x=0}^{s-1}\ket{x}\ket{0} \\
& \stackrel{U_{f}}{\longrightarrow} &\frac{1}{\sqrt{s}}\sum_{x=0}^{s-1}\ket{x}\ket{x + a^{x} \mbox{ mod $N$}} \label{eqn:Shorunitary}
\end{eqnarray}

Suppose $r$ divides $s$ exactly. A measurement of the output register in the computational basis would leave the input register in a state of the form:
\begin{equation}
\frac{1}{\sqrt{\frac{s}{r}}}\sum_{j=0}^{s/r - 1} \ket{x_{i} + jr} \label{eqn:inputstate}
\end{equation}
The value $x_{i}$ is the offset, which depends on the outcome $i$ of the measurement of the output register. The sum is taken over the values of $j$ for which $f(x_{i} + jr) = i$. Since the state label contains the random offset, a direct measurement of the label yields no information about the period.

A discrete quantum  Fourier transform for the integers mod $s$ is now applied to the input register, i.e., a unitary transformation:
\begin{equation}
\ket{x} \stackrel{U_{DFT_{s}}}{\longrightarrow} \frac{1}{\sqrt{s}}\sum_{y=0}^{s-1}e^{2\pi i \frac{xy}{s}}\ket{y}, \mbox{ for $x \in Z_{s}$} \label{eqn:Fourier}
\end{equation}
Note that  a Hadamard transformation is a discrete quantum Fourier transform for the integers mod 2, so this step is analogous to the application of the Hadamard transformation in Simon's algorithm. Under the Fourier transformation, the state of the input register undergoes the transition:
\begin{equation}
\frac{1}{\sqrt{\frac{s}{r}}}\sum_{j=0}^{\frac{s}{r} - 1} \ket{x_{i} + jr} \stackrel{U_{DFT_{s}}}{\longrightarrow}
\frac{1}{\sqrt{r}}\sum_{k=0}^{r-1}e^{2\pi i \frac{x_{i}k}{r}}\ket{ks/r} \label{eqn:Fourier2}
\end{equation}
where, similar to the derivation of (\ref{eqn:SimonHadamard}), the amplitudes are non-zero only if $y$ is not a multiple $k$ of $s/r$ (i.e., $\sum_{j=0}^{\frac{s}{r}- 1}e^{2\pi i\frac{jry}{s}} = s/r$ if $y = ks/r$; $\sum_{j=0}^{\frac{s}{r} - 1}e^{2\pi i\frac{jry}{s}} = 0$ if $y \neq ks/r$). The effect is to shift the offset into a phase factor and invert the period as a multiple of $s/r$. A measurement of the input register in the computational basis then yields $c = ks/r$. The algorithm is run a number of times until  a value of $k$ coprime to $r$ is obtained. Cancelling $c/s$ to lowest terms then yields $k$ and $r$ as $k/r$.

Suppose $r$ does not divide $s$ exactly. Then some of the states in (\ref{eqn:inputstate}) will have an additional term. For example, suppose $s = r + d$, where $d < r$. Then $d$ of the states in 
(\ref{eqn:inputstate}) will have an extra term and take the form:
\begin{equation}
\frac{1}{\sqrt{\frac{s}{r}}+1}\sum_{j=0}^{s/r} \ket{x_{i} + jr} 
\end{equation}
After the Fourier transformation, the expression (\ref{eqn:Fourier2}) will contain additional terms with negligible amplitudes for values of $s \neq k$ ($k = 0, 1, \ldots, r-1$) if $s/r$ is large.

Since the value of $r$ is unknown in advance of applying the algorithm, we do not, of course, recognize when a measurement outcome yields a value of $k$ coprime to $r$. The idea is to run the algorithm, cancel $c/s$ to lowest terms to obtain a candidate value for $r$ and hence a candidate factor of $N$, which can then be tested  by division into $N$. Even when we do obtain a value of $k$ coprime to $r$, some values of $a$ will yield a period for which the method fails to yield a factor of $N$, in which case we randomly choose a new value of $a$ and run the algorithm with this value. The point is that all these steps are efficient, i.e., can be performed in polynomial time, and since only a polynomial number of repetitions are required to determine a factor with any given probability $p < 1$, the algorithm is a polynomial-time algorithm, achieving an exponential speed-up over any known classical algorithm. 

To see how the algorithm works from a quantum logical perspective, consider the case $N= 15, a = 7$ and $s = 64$ discussed in \cite[p. 160]{Barenco}. In this case, the function $f(x) = a^{x}\mbox{ mod $15$}$ is:
\begin{eqnarray*}
7^{0} \mbox{ mod $15$} & = & 1\\
7^{1} \mbox{ mod $15$} & = & 7\\
7^{2} \mbox{ mod $15$} & = & 4\\
7^{3} \mbox{ mod $15$} & = & 13\\
7^{4} \mbox{ mod $15$} & = & 1\\
\vdots \\
7^{63} \mbox{ mod $15$} & = & 13
\end{eqnarray*}
and the period is evidently $r=4$, which divides $s=64$ exactly.\footnote{The factors 3 and 5 of 15 are derived as the greatest common factors of $a^{r/2} - 1 = 48$ and 15,  and $a^{r/2} + 1 = 50$ and 15, respectively.} After the application of the unitary transformation $U_{f} = a^{x}\mbox{ mod $N$}$, the state of the two registers is:
\begin{eqnarray}
& \frac{1}{8}(\ket{0}\ket{1} + \ket{1}\ket{7}  +  \ket{2}\ket{4} + \ket{3}\ket{13}  \nonumber \\
& \mbox{} + \ket{4}\ket{1} + \ket{5}\ket{7}  +  \ket{6}\ket{4} + \ket{7}\ket{13} \nonumber \\
& \vdots \nonumber \\
& \mbox{} + \ket{60}\ket{1} + \ket{61}\ket{7}  +  \ket{62}\ket{4} + \ket{63}\ket{13})
\end{eqnarray}
This is the state (\ref{eqn:Shorunitary}) for $s = 64$, $a = 7$. This state can be expressed as:
\begin{eqnarray}
& \frac{1}{4}(\ket{0} + \ket{4} + \ket{8} + \ldots + \ket{60})\ket{1} \nonumber \\
& + \frac{1}{4}(\ket{1} + \ket{5} + \ket{9} + \ldots + \ket{61})\ket{7} \nonumber \\
& + \frac{1}{4}(\ket{2} + \ket{6} + \ket{10} + \ldots + \ket{62})\ket{4} \nonumber \\
& + \frac{1}{4}(\ket{3} + \ket{7} + \ket{11} + \ldots + \ket{63})\ket{13})
\end{eqnarray}

A measurement of the output register would yield (equiprobably) one of four states for the input register, depending on the outcome of the measurement: 1, 7, 4, or 13:
\begin{eqnarray}
& \frac{1}{4}(\ket{0} + \ket{4} + \ket{8} + \ldots + \ket{60}) \label{eqn:1}\\
& \frac{1}{4}(\ket{1} + \ket{5} + \ket{9} + \ldots + \ket{61}) \label{eqn:2} \\
& \frac{1}{4}(\ket{2} + \ket{6} + \ket{10} + \ldots + \ket{62})  \\
& \frac{1}{4}(\ket{3} + \ket{7} + \ket{11} + \ldots + \ket{63}) 
\end{eqnarray}
These are the states (\ref{eqn:inputstate}) for values of the offset $x_{1} =0$, $x_{7} = 1$, $x_{4} = 2$, $x_{13} = 3$. 

Application of the quantum Fourier transform yields:
\begin{description} \centering
\item{$x_{1} = 0:$}  \mbox{ }$\frac{1}{2}(\ket{0} + \ket{16} + \ket{32}+\ket{48})$
\item{$x_{7} = 1:$} \mbox{ }$\frac{1}{2}(\ket{0} + i\ket{16} - \ket{32} - i\ket{48})$
\item{$x_{4} = 2:$} \mbox{ }$\frac{1}{2}(\ket{0} - \ket{16} + \ket{32} - \ket{48})$
\item{$x_{13} = 3:$} \mbox{ }$\frac{1}{2}(\ket{0} - i\ket{16} - \ket{32} + i\ket{48})$
\end{description}
which are the states in (\ref{eqn:Fourier2}). (Here $s = 64$, $r = 4$; $\sqrt{\frac{s}{r}} = 4$, $\frac{s}{r} - 1 = 15$.) So for the period $r = 4$, the state of the input register ends up in the 4-dimensional subspace spanned by the vectors $\ket{0}, \ket{16}, \ket{32}, \ket{48}$ making the corresponding quantum proposition true. 

Note that if, say, $s = 66$, so that the period $r = 4$ does not divide $s$ exactly, the input states relative to the output states $\ket{1}$ and $\ket{7}$ would each have an additional term, so the states (\ref{eqn:1}), (\ref{eqn:2}) would each have an additional term, $\ket{64}$ and 
$\ket{65}$, respectively. After the quantum Fourier transformation, the states for $x_{1} = 0$ and $x_{7} = 1$  would be a linear superposition of all the states $\ket{1}, \ldots, \ket{66}$, with small amplitudes for the states $\ket{i}, i \neq 0, 16, 32, 48$.

Now consider all possible even periods $r$ for which $f(x) = a^{x} \mbox{ mod $15$}$, where $a$ is coprime to $15$. The other possible values of $a$  are 2, 4, 8, 11, 13, 14 and the corresponding periods turn out to be 4, 2, 4, 2, 4, 2. So we need only consider $r = 2$.\footnote{Every value of $a$ except $a = 14$ yields the correct factors for 15. For $a = 14$, the method fails: $r = 2$, so $a^{\frac{r}{2}} = -1 \mbox{ mod $15$}$.} Note that different values of $a$ with the same period affect only the labels of the output register (e.g., for $a = 2$, the labels are $\ket{1}, \ket{2}, \ket{4}, \ket{8}$ instead of $\ket{1}, \ket{7}, \ket{4}, \ket{13}$). So different  $a$ values for the same period are irrelevant to the quantum algorithm.

For $r = 2$, if we measured the output register, we would obtain (equiprobably) one of two states for the input register, depending on the outcome of the measurement:
\begin{eqnarray}
& \ket{0} + \ket{2} + \ket{4} + \ldots + \ket{62} \\
& \ket{1} + \ket{3} + \ket{5} + \ldots + \ket{63}  
\end{eqnarray}
After the quantum Fourier transformation, these states are transformed to:
\begin{description} \centering
\item{$x_{a} = 0:$}  \mbox{ }$\ket{0} + \ket{32}$
\item{$x_{b} = 1:$} \mbox{ }$\ket{0}  - \ket{32}$
\end{description}

In this case, the 2-dimensional subspace $\hil{V}_{r=2}$ spanned by $\ket{0}, \ket{32}$ for $r = 2$ is included in the 4-dimensional subspace $\hil{V}_{r=4}$ for $r = 4$. A measurement can distinguish $r = 4$ from $r = 2$ reliably, i.e., whether the final state of the input register is in $\hil{V}_{r=4}$ or $\hil{V}_{r=2}$, only if the final state is in $\hil{V}_{r=4} - \hil{V}_{r=2}$, the part of $\hil{V}_{r=4}$ orthogonal to 
$\hil{V}_{r=2}$. What happens if the final state ends up in $\hil{V}_{r=2}$?

Shor's algorithm works as a randomized algorithm. As mentioned above, it produces a candidate value for the period $r$ and hence a candidate factor of $N$, which can be tested (in polynomial time) by division into $N$. A measurement of the input register in the computational basis yields an outcome $c = ks/r$. The value of $k$ is chosen equiprobably by the measurement of the output register. The procedure is to repeat the algorithm  until the outcome yields a value of $k$ coprime to $r$, in which case canceling $c/s$ to lowest terms yields $k$ and $r$ as $k/r$.

For example, suppose we choose $a = 7$, in which case (unknown to us) $r = 4$. The values of $k$ coprime to $r$ are $k = 1$ and  $k = 3$ (this is  also unknown to us, because $k$ depends on the value of $r$). Then $c/s$ cancelled to lowest terms is $1/4$ and $3/4$, respectively, both of which yield the correct period. From the geometrical perspective, these values of $k$ correspond to finding  the state after measurement in the computational basis to be $\ket{16}$ or $\ket{48}$, both of which do distinguish $\hil{V}_{r=4}$ from $\hil{V}_{r=2}$. 

Suppose we choose a value of $a$ with period $r = 2$ and find the value $c = 32$. The only value of $k$ coprime to $r$ is $k = 1$. Then $c/s$ cancelled to lowest terms is $1/2$, which yields the correct period, and hence the correct factors of $N$. But  $c = 32$ could also be obtained for $a = 7$, $r = 4$, and $k = 2$, which does not yield the correct period, and hence does not yield the correct factors of $N$. Putting it geometrically: the value $k = 1$ for $r = 2$ corresponds to the same state, $\ket{32}$, as the value $k = 2$ for $r = 4$. Once we obtain the candidate period $r = 2$ (by cancelling $c/s = 32/64$ to lowest terms), we calculate the factors of $N$ as the greatest common factors of $a \pm 1$ and $N$ and test these by division into $N$. If $a = 7$, these calculated factors will be incorrect. If $a = 2$, say, the factors calculated in this way will be correct.

\section{Conclusion}

Simon's algorithm and Shor's algorithm work as period-finding algorithms by encoding alternative partitions of the domain of a function, defined by alternative possible periods, as quantum propositions represented by subspaces in a Hilbert space, which are orthogonal except for overlaps. The subspace corresponding to a particular partition is spanned by orthogonal linear superpositions of states associated with the elements in  the (mutually exclusive and collectively exhaustive) subsets of the partition. The period-finding algorithm is designed to produce an entangled state in which such superpositions, representing states of an input register, are correlated with distinct orthogonal states of an output register. The reduced state of the input register is then an equal-weight mixture of states spanning the subspace corresponding to the partition, where each state encodes a subset in the partition as a linear superposition of the elements in the subset. Since the subspaces are represented by commuting projection operators, a measurement of the state of the  input register in a certain basis can reveal the subspace containing the state, and hence the period associated with the partition, except when the measurement projects the state onto the overlap region. This measurement basis is unitarily related to the computational basis by a known unitary transformation that can be implemented efficiently, so a measurement in the computational basis after this unitary information will yield the same information. This is the function of the final Hadamard transformation or discrete quantum Fourier transformation, and the possibility of an efficient implementation of this transformation is crucial to the efficiency of the algorithm. By contrast with the classical `fast Fourier transform,' it is a remarkable feature of the discrete quantum Fourier transform that it can be implemented efficiently.

The information-processing in Deutsch's XOR algorithm has a similar quantum logical interpretation in terms of the subspace structure of Hilbert space. The problem here is to distinguish two alternative disjunctive properties of a function  ($0 \rightarrow 0$ and $1 \rightarrow 0$ \emph{or} $0 \rightarrow 1$ and $1 \rightarrow 1$ for a constant function, versus   $0 \rightarrow 0$ and $1 \rightarrow 1$ \emph{or} $0 \rightarrow 1$ and $1 \rightarrow 0$ for a balanced function), which are encoded as two planes in a 4-dimensional Hilbert space (orthogonal except for an overlap). Each disjunct in the disjunction is a conjunction of two elements (e.g., $0 \rightarrow 0$ \emph{and} $1 \rightarrow 0$). The plane corresponding to a particular disjunction is spanned by a pair of states that encode the elements of the conjunctions as linear superpositions. The algorithm is designed to produce one of these states, depending on which disjunction is true of the function. From this perspective, the XOR algorithm appears directly as a special case of Simon's algorithm.

The first stage of a quantum algorithm designed to evaluate some global property of a function involves the creation of an entangled state of the input and output registers in which every value in the domain of the function  is correlated with a corresponding value in its range. This is referred to as `quantum parallelism' and is often cited as the source of the speed-up in a quantum computation. The idea is that a quantum computation is something like a massively parallel classical computation, for all possible values of a function. This appears to be Deutsch's view \cite{Deutsch97}: in an Everettian many-worlds interpretation of quantum mechanics,  the parallel computations can be regarded as taking place in parallel  universes. (For a critique, see \cite{Steane03}.) 

From the quantum logical perspective outlined here, the picture is entirely different. Rather than `computing all values of a function at once,' a quantum algorithm achieves an exponential speed-up over a classical algorithm precisely by avoiding the computation of \emph{any} values of the function at all. This is redundant information for a quantum algorithm but essential information for a classical algorithm. The trick in a quantum algorithm is to exploit the non-Boolean logic represented by the projective geometry of Hilbert space to encode a global property of a function (such as a period, or a disjunctive property) as a subspace in Hilbert space, which can be efficiently distinguished from alternative subspaces corresponding to alternative global properties by a measurement that determines the target subspace as the subspace containing the final state produced by the algorithm. The point of the procedure is precisely to avoid the evaluation of the function in the determination of the global property, in the sense of producing a value in the range of the function for a value in its domain, and it is this feature---impossible in the Boolean logic of classical computation---that leads to the speed-up relative to classical algorithms.

\section*{Acknowledgments}

Support for research leading to this paper is acknowledged from the  University of Maryland General Research Board (2005), the National Science Foundation (2006), and the Perimeter Institute for Theoretical Physics in Waterloo, Canada, where the paper was written during a stay as a long-term visiting researcher in 2006. I thank Richard Jozsa for illuminating correspondence, and Hans Briegel for suggesting that it would be interesting to look at quantum computation from a quantum logical perspective.

\end{document}